\documentclass{article}
\usepackage{spconf,amsmath,graphicx,hyperref}
\usepackage{subcaption}
\usepackage{amssymb}
\usepackage{multirow}
\usepackage{cite}
\usepackage{comment}

\title{Predicting Music Valence and Arousal via Textual and Multimodal In-Context Learning in Large Language Models}
\name{
Chi-En Dai,
Chih-Pin Tan,
Yi-Hsuan Yang
}

\address{
Graduate Institute of Communication Engineering, National Taiwan University, Taiwan\\
}

\begin{document}
%
\maketitle
\begin{abstract}

While arousal can be reliably predicted from musical audio, emotional valence remains difficult to model due to its subjective and context-dependent nature. In this work, we investigate the potential of in-context learning (ICL) with large language models to advance continuous valence and arousal prediction. Rather than predicting emotion scores directly, our framework leverages contextual demonstrations from an auxiliary dataset to refine the initial predictions produced by a small, task-specific base model. We systematically evaluate a text-based ICL variant using lyrics and generated audio captions, alongside a multimodal variant that  incorporates raw audio signals. Using the entire MER60 dataset strictly as a held-out test set, our experiments demonstrate substantial performance gains, particularly for valence. Text-based ICL improves the valence $R^2$ from 0.110 to 0.427 with 80 demonstration shots, while the multimodal variant achieves 0.346 with only 10 shots. These findings suggest that training-free ICL offers an effective inference-time strategy for current language models to predict continuous music emotion.

\end{abstract}
\begin{keywords}
Music emotion recognition
\end{keywords}

\section{Introduction}\label{sec:intro}

Music emotion recognition (MER) aims to estimate the perceived emotional response of listeners \cite{yang_mer2012,kang2025mer}. Emotions are typically modeled using categorical labels or continuous dimensions. In this work, we focus on the continuous valence--arousal framework \cite{russell1980, circumplex2005}, where \textit{arousal} indicates emotional energy and \textit{valence} measures emotional positivity or negativity. While numerous classical \cite{yang_regression2008,Hu_CH818_2017} and deep learning methods \cite{hershey2017cnn, Music2Emo2025} have been proposed to predict valence and arousal from musical audio, valence prediction remains challenging due to its highly abstract and context-dependent nature \cite{yang_regression2008, multimodal2018}.

Recently, large language models (LLMs) and their multimodal counterparts (MLLMs) have demonstrated remarkable capabilities across diverse tasks \cite{wei2022cot, rubenstein2023audiopalm, tang2023salmonn, qwen3omni2025}. However, their direct zero-shot performance in continuous valence--arousal prediction from musical audio remains largely unexplored. In our pilot study, we employed MER60 \cite{MER602011}, a benchmark dataset of 60 Western pop songs compiled before 2011, to test two representative models: GPT-5 (a text-based LLM) and Qwen3-Omni 30B \cite{qwen3omni2025} (an MLLM). Evaluating with the coefficient of determination ($R^2$), where 1.0 indicates perfect prediction and negative values signal severe performance degradation \cite{R2_1998}, GPT-5 achieved $R^2_A = 0.699$ for arousal but $R^2_V = -0.801$ for valence. Similarly, Qwen3-Omni 30B obtained $R^2_A = 0.764$ 
and $R^2_V = 0.037$. 
While both models perform competitively on arousal—approaching established baselines in the literature (e.g., $R^2_A = 0.800$ reported in \cite{Hu_CH818_2017})—their zero-shot valence predictions fail dramatically, falling far below prior audio-based benchmarks (e.g., $R^2_V = 0.270$ in \cite{Hu_CH818_2017}). Crucially, because the songs in MER60 have been released for over 15 years, likely the text descriptions or audio clips related to them were already included in the massive pre-training corpora of the LLM/MLLM. The fact that zero-shot valence prediction remains  poor indicates that potential song exposure or text-based memorization alone is insufficient for zero-shot valence reasoning, highlighting the need for explicit, context-guided alignment.

The goal of this work is therefore to study whether we can effectively improve the ability of current LLMs and MLLMs to predict valence from music. More specifically, since training such large-scale models from scratch is computationally prohibitive, we ask whether we can leverage the prominent in-context learning (ICL) paradigm \cite{brown2020gpt3, manyshot_icl2024}. By exposing pre-trained LLMs and MLLMs to contextual demonstrations consisting of music audio and emotion labels from an auxiliary dataset (specifically, CH818 \cite{CH8182017}) that is completely disjoint from our test set (i.e., MER60), we aim to boost their valence prediction performance. Because this ICL approach operates entirely at inference time and requires no weight updates or task-specific fine-tuning \cite{brown2020gpt3}, it provides a straightforward, flexible, and practical solution for MER.

\begin{figure*}[t]
    \centering
    \includegraphics[width=0.75\textwidth]{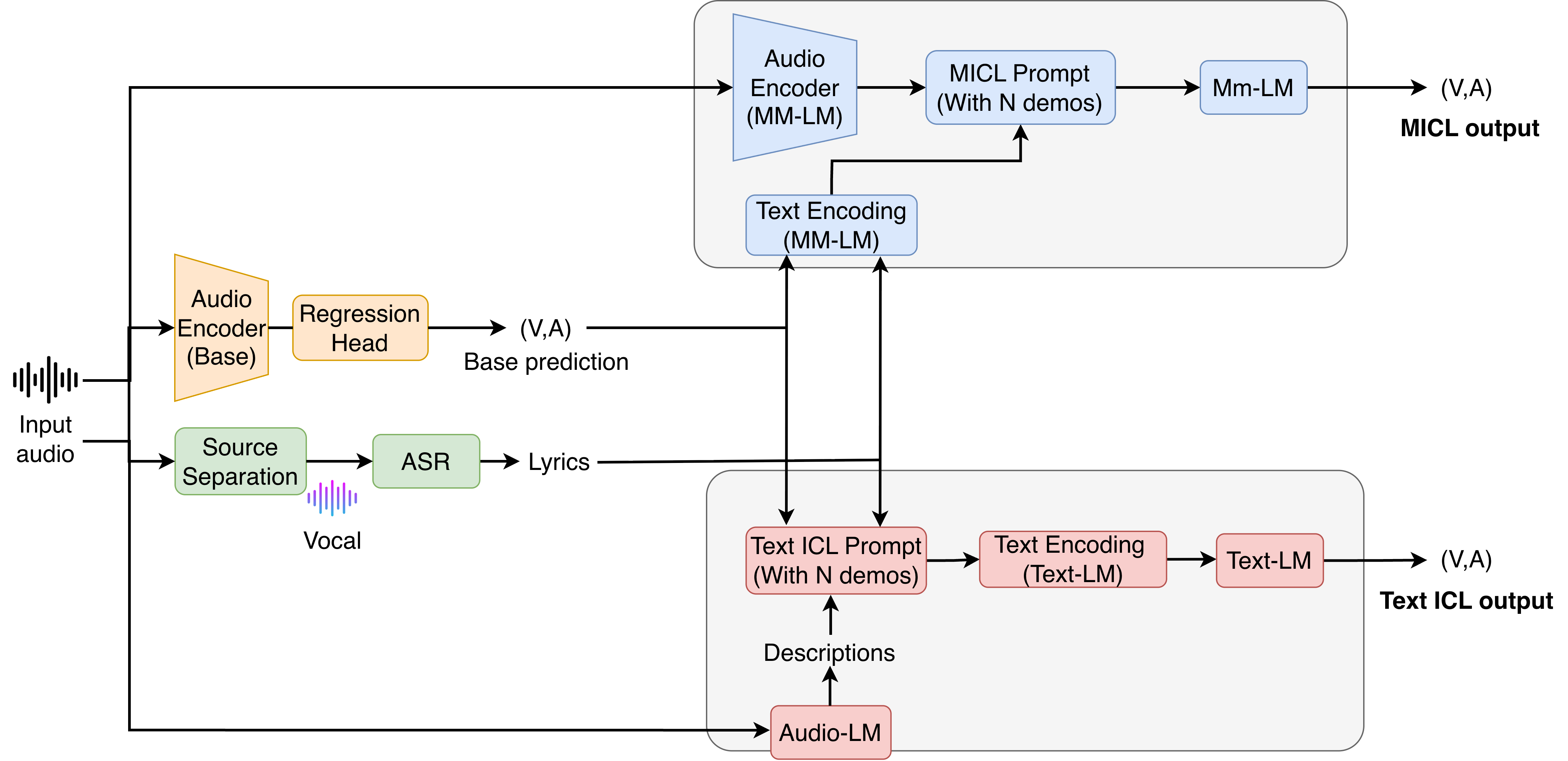}
    \caption{Overview of the proposed framework. A base audio model produces initial valence–arousal (VA) predictions, 
    which are refined via in-context learning using either a text-only LLM (bottom) or a multimodal LLM (top).}
    \label{fig:architecture}
\end{figure*}



Specifically, text-only LLMs cannot process raw audio signals directly, necessitating text-based ``surrogates'' to represent the musical content of demonstration songs during ICL. Therefore, as our first research question, we evaluate the effectiveness of song lyrics versus automatic audio captions as  surrogates. We consider lyrics under the assumption that they convey rich semantic and thematic context essential for valence perception. Similarly, for audio captions, large audio-language models (e.g., Qwen-Audio \cite{qwen2audio}) can be prompted to extract high-level descriptions such as emotional atmosphere and mood dynamics. Furthermore, while MLLMs can ingest raw audio signals directly, we similarly evaluate whether augmenting the audio input with lyrics 
enhances valence prediction in a multimodal ICL setting.


As our second research question, we study whether incorporating a \emph{small}, task-specific deep learning model trained explicitly for valence prediction can further improve result. 
The core idea is that a compact base model can collaborate with the \emph{large} language models  by simplifying the in-context learning task. 
Specifically, instead of asking the language model to predict absolute valence values from scratch, we use the base model's initial prediction as an anchor and prompt the LLM/MLLM to estimate a residual offset. This predicted delta is then added to the base estimate to refine the final emotion scores. 
This base model is trained using the entire auxiliary dataset CH818,
ensuring that the overall framework maintains strict cross-dataset generalization.


To provide further insights, we systematically investigate key factors such as the required number of demonstration shots and the impact of LLM/MLLM model capacity. Furthermore, we conduct all experiments for both valence and arousal to examine whether the observed gains are dimension-specific. Our findings show that ICL can indeed greatly improve the valence prediction capabilities of current LLMs and MLLMs, though current text-based and multimodal architectures exhibit different optimal configurations.

\section{Related Work}\label{sec:related work}


MER research can be categorized in several ways, according to the emotion model (categorical or continuous emotions \cite{yang_mer2012}), the nature of affect (perceived or felt emotion \cite{koelstra2011deap}), or the input modality (e.g., audio signals, lyrics \cite{multimodal2018}, or symbolic representations like MIDI \cite{hung2021emopia}). 
In recent years, deep learning architectures incorporating convolutional/recurrent networks and pre-trained backbones have been studied for various MER settings \cite{hung2021emopia,cnn_bilstm2025, Music2Emo2025}. 
Here, we focus specifically on continuous, perceived MER from audio clips. 

The annotation quality of training datasets strongly affects the generalizability of MER models \cite{Hu_CH818_2017}. Recent audio-based continuous MER studies mostly rely on the DEAM dataset \cite{DEAM2017}, largely due to its public accessibility. 
However, DEAM was annotated via crowdsourcing,
leading to variable label quality.
In contrast, older datasets such as MER60 \cite{MER602011} and CH818 \cite{CH8182017} were collected with 
expert-supervised annotation protocols, yielding substantially higher reliability, though restricted distribution has historically limited their usage. 
As the creators of CH818 and MER60, we leverage both in this study. For evaluation integrity, we utilize CH818 (covering 818 Chinese pop clips with medium-to-high annotation rigor) as our auxiliary dataset for base model training and ICL demonstrations, while using the entire MER60  (60 Western pop with high annotation rigor) exclusively as the test set.


Recent advances in LLMs/MLLMs have spurred growing interest in applying foundation models to music \cite{qwen2audio, qwen3omni2025}. 
However, their application to MER remains nascent. To date, existing work such as Yang et al. \cite{llm_annotation2025} primarily explores LLMs for categorical emotion annotation for symbolic MIDI.
In contrast, our work targets the more challenging task of continuous valence--arousal regression directly from musical audio. Furthermore, while ICL has proven highly effective at enabling LLMs to perform complex reasoning via demonstrations  \cite{brown2020gpt3, manyshot_icl2024}, 
its potential for continuous regression in audio and music remains largely unexplored. 

\begin{figure}[t]
\centering

\begin{subfigure}{0.45\linewidth}
    \raggedright
    \footnotesize
    \fbox{
    \begin{minipage}{0.80\linewidth}

    \textbf{Header:} 
    Task Definition \\

    \textbf{Demonstration 1:} \\
    Description: ... \\
    Lyrics: ... \\
    Prior: $(v_\text{base}, a_\text{base})$ \\
    GT: $(v_\text{gt}, a_\text{gt})$ \\
    Delta: $(\Delta v, \Delta a)$\\
    \textbf{...} \\

    \textbf{Demonstration $N$:} \\
    Description: ... \\
    Lyrics: ... \\
    Prior: $(v_\text{base}, a_\text{base})$ \\
    GT: $(v_\text{gt}, a_\text{gt})$ \\
    Delta: $(\Delta v, \Delta a)$\\
    
    \textbf{Query:} \\
    Description: ... \\
    Lyrics: ... \\
    Prior: $(v_\text{base}, a_\text{base})$ \\

    \textbf{Output:} \\
    $(\Delta v, \Delta a)$

    \end{minipage}
    }
    \caption{Text-based ICL}
\end{subfigure}
\hspace{0.02\linewidth}
\begin{subfigure}{0.45\linewidth}
    \raggedright
    \footnotesize
    \fbox{
    \begin{minipage}{0.80\linewidth}

    \textbf{Header:} 
    Task Definition \\

    \textbf{Demonstration 1:} \\
    Audio: \textit{$<$audio emb$>$} \\
    Lyrics: ... \\
    Prior: $(v_{\mathrm{prior}}, a_{\mathrm{prior}})$ \\
    GT: $(v_{\mathrm{gt}}, a_{\mathrm{gt}})$ \\
    Delta: $(\Delta v, \Delta a)$\\
    \textbf{...} \\
    
    \textbf{Demonstration $N$:} \\
    Audio: \textit{$<$audio emb$>$} \\
    Lyrics: ... \\
    Prior: $(v_{\mathrm{prior}}, a_{\mathrm{prior}})$ \\
    GT: $(v_{\mathrm{gt}}, a_{\mathrm{gt}})$ \\
    Delta: $(\Delta v, \Delta a)$\\
    
    \textbf{Query:} \\
    Audio: \textit{$<$audio emb$>$} \\
    Lyrics: ... \\
    Prior: $(v_{\mathrm{prior}}, a_{\mathrm{prior}})$ \\

    \textbf{Output:} \\
    $(\Delta v, \Delta a)$

    \end{minipage}
    }
    \caption{MICL}
\end{subfigure}
\caption{Prompt structure of text and multimodal ICL, where $(v_\text{base}, a_\text{base})=\mathbf{b}$ denotes the prediction of the base model, and $(v_\text{gt}, a_\text{gt})=\mathbf{y}$ the ground-truth VA values for a clip.}
\label{fig:icl_prompt}
\end{figure}

\begin{table*}[!t]
\centering
\begin{tabular}{lrlrrrrrr}
\hline
\multirow{2}{*}{Model} & \multirow{2}{*}{Size} & \multirow{2}{*}{Training Data} 
& \multicolumn{2}{c}{DEAM \cite{DEAM2017}} 
& \multicolumn{2}{c}{CH818 \cite{CH8182017}} 
& \multicolumn{2}{c}{MER60 \cite{MER602011}} \\
 & 
& 
& $R^2_V$ & $R^2_A$ 
& $R^2_V$ & $R^2_A$ 
& $R^2_V$ & $R^2_A$ \\
\hline
SVR-based \cite{Hu_CH818_2017}  & $<$1M         & CH818                         & N/A & N/A & 0.250 & 0.770 & 0.270 & 0.800\\
VGGish \cite{essentia2013}      & 62M     & DEAM                          & 0.520 & 0.611 & 0.122 & 0.256 & --0.071 & 0.424 \\
Music2Emo \cite{Music2Emo2025}  & 100M    & DEAM$+$EmoMusic$+$PMemo       & 0.518 & 0.623 & 0.204 & 0.276 & --0.438 & 0.511 \\
\hline
Base model (ours)               & 91M           & DEAM                          & 0.542 & 0.529 & --0.024 & 0.099 & --0.127 & 0.660 \\
Base model (ours)               & 91M           & CH818                         & --0.062 & 0.168 & 0.430 & 0.782 & 0.110 & 0.783 \\
\hline
\end{tabular}
\caption{Comparison of baseline models under within- and cross-dataset settings.}
\label{tab:baseline_compare}
\end{table*}

\section{Methodology}\label{sec:methodology}


Our framework consists of two stages, as illustrated in Fig.~\ref{fig:architecture}. Given an input audio $\mathbf{x}$, a base model $f(\cdot)$ first predicts an initial valence--arousal (VA) estimate $\mathbf{b}=f(\mathbf{x})\in[1,9]^2$, where valence and arousal scores are normalized to the standard 1--9 scale. A second-stage LLM or MLLM $g(\cdot)$ then predicts a calibration offset $\Delta$, yielding the final prediction $\hat{\mathbf{y}}=\mathbf{b}+\Delta$, which is clipped to the valid VA range. 
Predicting the residual $\Delta$ instead of absolute VA values allows the second-stage model to focus only on correcting the base prediction.

In our implementation, the \textbf{base model} follows the audio branch of Music2Emo \cite{Music2Emo2025}, replacing MERT \cite{li2024mert} with BEATs embeddings \cite{BEATs2022}. For each  clip, 5-second BEATs features are averaged into a 768-dimensional representation and fed into an MLP with batch normalization, ReLU, and dropout. The model is trained using AdamW (learning rate $10^{-4}$), batch size 64, and mean squared error (MSE) loss for 150 epochs.


As the zero-shot performance of current LLMs or MLLMs for VA prediction is unsatisfactory, we employ ICL by pre-pending $N$ demonstrations at inference time 
for each testing audio clip (i.e., the ``query''), where $N$ denotes the number of shots. 
We investigate two variants in this work. 

For \textbf{text-based ICL}, as shown in Fig.~\ref{fig:icl_prompt}(a), each demonstration consists of the base prediction $\mathbf{b}$, lyrics $\ell$, audio description $s$, ground-truth VA values $\mathbf{y}$, and the target offset $\Delta=\mathbf{y}-\mathbf{b}$. The $N$ demonstrations and the query are serialized into a prompt, enabling the LLM to predict the calibration offset $\Delta=g^{(\text{text-ICL})}(\mathbf{b},\ell,s)$ for the testing clip. 


To obtain rich textual surrogates for the musical content, lyrics $\ell$ are transcribed using Whisper-large-v3 \cite{radford2023whisper} on vocal tracks isolated via HTDemucs source separation \cite{defossez2021hybrid}. For audio descriptions $s$, we prompt Qwen2-Audio \cite{qwen2audio} to generate one coherent text paragraph in English for each input audio clip that reflect ``the listener’s feelings'' covering the following aspects: i) overall emotion or mood of the song, ii) energy level and tempo feeling, iii) emotional color of melody and harmony, and iv) vocal or instrumental expression. 

For \textbf{multimodal ICL} (MICL), as shown in Fig.~\ref{fig:icl_prompt}(b), the multimodal variant replaces the audio description with the original audio while maintaining the same underlying prompt structure, yielding $\Delta=g^{(\text{MICL})}(\mathbf{b},\ell,\mathbf{x})$. 
Audio clips are processed directly by the MLLM's native audio encoder, interleaving audio tokens with text in the context window.
This way, the model utilizes both acoustic and textual information.

\section{Experiment}
\label{sec:exp}


We use CH818 \cite{CH8182017} for base model training and ICL demonstrations, and MER60 \cite{MER602011} as the held-out test set. CH818 consists of 818 clips (30s each) annotated by 3 dedicated music experts, providing consistent medium-to-high annotation reliability. MER60 comprises 60 clips (also 30s each), each labeled by 40 human subjects in a controlled laboratory setup using a 2D emotional plane interface, providing ground truth labels with high annotation rigor ideal for evaluating ICL refinement. For each clip, we use the average VA values across annotators as the ground truth. 






\subsection{Evaluation of the Base Model}\label{sec:exp:base}

We compare our base model with three existing ones, SVR \cite{Hu_CH818_2017}, VGGish \cite{essentia2013}, and Music2Emo \cite{Music2Emo2025}, trained with different datasets. 
As shown in Table~\ref{tab:baseline_compare}, while all models perform well on the validation split of their native training data, they generalize poorly to the out-of-domain MER60, particularly for valence prediction. 
Our CH818-trained base model achieves the best cross-dataset performance among the deep learning models while remaining competitive with the classical SVR baseline, motivating its usage as the foundational anchor.


\begin{table}[!t]
    \centering
    \begin{subtable}{\linewidth}
    \centering
    \begin{tabular}{c|cc}
    \hline
    Number of shots & $R^2_V$ & $R^2_A$ \\
    \hline
    ~~0-shot  & --0.801~~~~~~~~~~~~~ & 0.699~~~~~~~~~~~~~ \\
    ~~5-shot  & ~~0.188 \small{$\pm$ 0.036} & 0.805 \small{$\pm$ 0.014} \\
    10-shot   & ~~0.274 \small{$\pm$ 0.054} & 0.858 \small{$\pm$ 0.015} \\
    20-shot   & ~~0.271 \small{$\pm$ 0.025} & 0.867 \small{$\pm$ 0.010} \\  
    40-shot   & ~~0.354 \small{$\pm$ 0.036} & 0.870 \small{$\pm$ 0.013} \\
    80-shot   & ~~\textbf{0.427} \small{$\pm$ 0.054} & \textbf{0.874} \small{$\pm$ 0.010} \\
    \hline
    \end{tabular}
    \caption{Scaling effect of text-based ICL}
    \end{subtable}
    
    \vspace{6pt}
    
    \begin{subtable}{\linewidth}
    \centering
    \begin{tabular}{c|ccc|cc}
    \hline
    ID &Base VA & Description & Lyric & $R^2_V$ & $R^2_A$ \\
    \hline
    1&$\surd$ &   &                & ~~0.002 & 0.535 \\
    2&  & $\surd$ &            & ~~0.209 & 0.676 \\
    3&  &   &  $\surd$            & --0.271 & 0.085 \\
    \hline
    4&$\surd$  & $\surd$  &          & ~~0.306 & 0.655 \\
    5&$\surd$  &   &  $\surd$            & ~~0.137 & 0.259 \\
    6&  &  $\surd$ &  $\surd$       & ~~0.104 & 0.671 \\
    \hline  
    7&$\surd$  & $\surd$  & $\surd$   & ~~\textbf{0.427} & \textbf{0.874} \\
    \hline
    \end{tabular}
    \caption{Ablation study of input components under 80-shot}
    \end{subtable}
    
    \caption{Text-based ICL results (mean±STD over five runs).}
    \label{tab:text_icl}
\end{table}

\subsection{Evaluation of Text-based ICL and MICL}\label{exp:text-ICL}


We use GPT-5 as the LLM for text-based ICL, and Qwen3-Omni 30B \cite{qwen3omni2025} for MICL.
All inference runs adopt a decoding temperature of 0, and reported results are averaged over 5 independent runs with randomly sampled demonstrations.


Table~\ref{tab:text_icl}(a) shows the effect of the number of demonstrations for text-based ICL. Without ICL, GPT-5 obtains a much lower $R^2_V$ than the base model. Introducing only five shots greatly improves both $R^2_V$ and $R^2_A$, while increasing the number of shots further benefits $R^2_V$. This indicates that contextual demonstrations effectively guide refinement, and that larger in-context sets provide richer information for valence.

Table~\ref{tab:text_icl}(b) reports the corresponding input ablation under $N=80$ (i.e., the best setting here). Using the base prediction alone does not work well, while removing the base prediction also degrades performance, validating the proposed two-stage framework. Among individual inputs, the audio description is the most informative, whereas lyrics alone contribute little. The best performance is achieved by combining all three inputs, suggesting that the base prediction provides a stable anchor, the audio description supplies the primary contextual information, and the lyrics offer complementary cues.

\begin{table}[!t]
    \centering
    \begin{subtable}{\linewidth}
    \centering
    \begin{tabular}{c|cc}
    \hline
    Number of shots & $R^2_V$ & $R^2_A$ \\
    \hline
    ~~0-shot  & ~~0.037~~~~~~~~~~~~~  & 0.764~~~~~~~~~~~~~ \\
    ~~5-shot  & ~~0.283 {\small{$\pm$ 0.109}} & 0.730 {\small{$\pm$ 0.008}} \\
    10-shot   & ~~\textbf{0.346} {\small{$\pm$ 0.046}} & 0.768 {\small{$\pm$ 0.018}} \\
    20-shot   & ~~0.247 {\small{$\pm$ 0.041}} & 0.782 {\small{$\pm$ 0.013}} \\
    40-shot   & ~~0.245 {\small{$\pm$ 0.031}} & \textbf{0.786} {\small{$\pm$ 0.041}} \\
    80-shot   & ~~0.238 {\small{$\pm$ 0.051}} & 0.783 {\small{$\pm$ 0.026}} \\
    \hline
    \end{tabular}
    \caption{Scaling effect of multimodal ICL}
    \end{subtable}
    
    \vspace{6pt}
    
    \begin{subtable}{\linewidth}
    \centering
    \begin{tabular}{c|ccc|cc}
    \hline
    ID & Base VA & Description & Lyric & $R^2_V$ & $R^2_A$ \\
    \hline
    1 & $\surd$ &   &                   & ~~0.176 & \textbf{0.796} \\
    2 & & $\surd$ &                     & --0.487 & 0.760 \\
    3 & &   &  $\surd$                  & --0.339 & 0.663 \\
    \hline
    4 & $\surd$  & $\surd$  &           & ~~0.251 & 0.773 \\
    5 & $\surd$  &   &  $\surd$         & ~~\textbf{0.346} & 0.768 \\
    6 &   &  $\surd$ &  $\surd$         & --0.321 & 0.766 \\
    \hline  
    7 & $\surd$  & $\surd$  & $\surd$   & ~~0.263 & 0.760 \\
    \hline
    \end{tabular}
    \caption{Ablation study of input components under 10-shot}
    \end{subtable}
    
    \caption{MICL results (mean±STD over five runs).}
    \label{tab:audio_text_icl}
\end{table}


Table~\ref{tab:audio_text_icl} reports the result for MICL.
Similar to text-based ICL, the MLLM performs poorly without demonstrations but improves substantially with only five shots, mainly for valence. Unlike text-based ICL, however, performance peaks at 10 shots and slightly degrades thereafter. We conjecture that the increasing context length caused by audio inputs limits the effectiveness of incorporating more demonstrations.

Table~\ref{tab:audio_text_icl}(b) presents the input ablation under $N=10$. 
Unlike text-based ICL, the base prediction plays a more dominant role in MICL, as using it alone already achieves competitive performance, further supporting the two-stage  framework. The best result is obtained by combining the base prediction with lyrics (ID~5). Adding audio descriptions 
slightly hurts $R^2_V$, 
suggesting that raw audio provides sufficient acoustic information, rendering audio descriptions redundant.

Finally, 
for 80-shot text-based ICL, replacing GPT-5 by Qwen2-7B, Qwen3.5-122B, and GPT-5-mini reduces  $R^2_V$ to 0.164, 0.285, 0.317, respectively. 
For 10-shot MICL, replacing Qwen3-Omni-30B by Qwen2.5-Omni-3B and Qwen2.5-Omni-7B  reduces  $R^2_V$ to 0.123 and 0.308. 
This suggests that larger models lead to better result in general.

\vspace{-3pt}



\vspace{1mm}
\section{Conclusion}

We proposed an 
ICL approach for dimensional MER,
leveraging textual and multimodal demonstrations to
predict  offsets that refine a base model prediction. 
Experiments on the high-quality but relatively small MER60 dataset demonstrate the effectiveness of this approach.
Future work will seek ways to  improve the multimodal ICL and expand the test data further. 
Code, model checkpoint, qualitative interpretable reasoning examples, prediction results on non-MER60 audio clips, and extension to full-song prediction are available at ~\url{https://github.com/DCN2001/MusicEmotionModeling}. 


\clearpage

\bibliographystyle{IEEEbib}
\bibliography{strings,refs}

@inproceedings{wei2022cot,
  author    = {Wei, Jason and others},
  title     = {Chain-of-Thought Prompting Elicits Reasoning in Large Language Models},
  booktitle = {Proc. NeurIPS},
  year      = {2022}
}

@article{rubenstein2023audiopalm,
  author  = {Rubenstein, Paul K. and others},
  title   = {{AudioPaLM}: A Large Language Model That Can Speak and Listen},
  journal = {arXiv preprint arXiv:2306.12925},
  year    = {2023}
}

@article{tang2023salmonn,
  author  = {Tang, Changli and others},
  title   = {{SALMONN}: Towards Generic Hearing Abilities for Large Language Models},
  journal = {arXiv preprint arXiv:2310.13289},
  year    = {2023}
}

@article{yang_mer2012,
  author  = {Yang, Yi-Hsuan and Chen, Homer H.},
  title   = {Machine Recognition of Music Emotion: A Review},
  journal = {ACM Trans. Intell. Syst. Technol.},
  volume  = {3},
  number  = {3},
  year    = {2012}
}

@article{kang2025mer,
  author  = {Kang, Jaeyong and Herremans, Dorien},
  title   = {Are We There Yet? {A} Brief Survey of Music Emotion Prediction Datasets, Models, and Outstanding Challenges},
  journal = {IEEE Trans. Affect. Comput.},
  year    = {2025}
}

@article{russell1980,
  author  = {Russell, James A.},
  title   = {A Circumplex Model of Affect},
  journal = {J. Pers. Soc. Psychol.},
  volume  = {39},
  number  = {6},
  pages   = {1161--1178},
  year    = {1980}
}

@article{circumplex2005,
  author  = {Posner, Jonathan and Russell, James A. and Peterson, Bradley S.},
  title   = {The Circumplex Model of Affect: An Integrative Approach to Affective Neuroscience, Cognitive Development, and Psychopathology},
  journal = {Dev. Psychopathol.},
  volume  = {17},
  number  = {3},
  pages   = {715--734},
  year    = {2005}
}

@inproceedings{Music2Emo2025,
  author  = {Kang, Jaeyong and Herremans, Dorien},
  title   = {Towards Unified Music Emotion Recognition Across Dimensional and Categorical Models},
  booktitle = {Proc. ICPR},
  year    = {2026}
}

@inproceedings{hershey2017cnn,
  author    = {Hershey, Shawn and others},
  title     = {{CNN} Architectures for Large-Scale Audio Classification},
  booktitle = {Proc. ICASSP},
  year      = {2017}
}

@inproceedings{essentia2013,
  author    = {Bogdanov, Dmitry and others},
  title     = {{Essentia}: An Audio Analysis Library for Music Information Retrieval},
  booktitle = {Proc. ISMIR},
  year      = {2013}
}

@inproceedings{hung2021emopia,
  author    = {Hung, Hsiao-Tzu and others},
  title     = {{EMOPIA}: A Multi-Modal Pop Piano Dataset For Emotion Recognition and Emotion-Based Music Generation},
  booktitle = {Proc. ISMIR},
  year      = {2021}
}

@article{koelstra2011deap,
  author    = {Koelstra, Sander and others},
  title     = {{DEAP}: A Database for Emotion Analysis Using Physiological Signals},
  journal   = {IEEE Trans. Affect. Comput.},
  volume    = {3},
  number    = {1},
  pages     = {18--31},
  year      = {2012},
}

@article{DEAM2017,
  author  = {Aljanaki, Anna and others},
  title   = {Developing a Benchmark for Emotional Analysis of Music},
  journal = {PLOS ONE},
  volume  = {12},
  number  = {3},
  pages   = {e0173392},
  year    = {2017}
}

@article{CH8182017,
  author  = {Hu, Xiao and Yang, Yi-Hsuan},
  title   = {The Mood of {Chinese} Pop Music: {Representation} and Recognition},
  journal = {J. Assoc. Inf. Sci. Technol.},
  volume  = {68},
  number  = {8},
  pages   = {1899--1910},
  year    = {2017}
}

@article{Hu_CH818_2017,
  author    = {Hu, Xiao and Yang, Yi-Hsuan},
  title     = {Cross-Dataset and Cross-Cultural Music Mood Prediction: {A} Case on {Western} and {Chinese} Pop Songs},
  journal   = {IEEE Trans. Affect. Comput.},
  volume    = {8},
  number    = {2},
  pages     = {228--240},
  year      = {2017},
}

@article{MER602011,
  author  = {Yang, Yi-Hsuan and Chen, Homer H.},
  title   = {Predicting the Distribution of Perceived Emotions of a Music Signal for Content Retrieval},
  journal = {IEEE Trans. Audio Speech Lang. Process.},
  volume  = {19},
  number  = {7},
  pages   = {2184--2196},
  month   = sep,
  year    = {2011}
}

@article{yang_regression2008,
  author  = {Yang, Yi-Hsuan and others},
  title   = {A Regression Approach to Music Emotion Recognition},
  journal = {IEEE Trans. Audio Speech Lang. Process.},
  volume  = {16},
  number  = {2},
  pages   = {448--457},
  month   = feb,
  year    = {2008}
}

@article{cnn_bilstm2025,
  author  = {Qiu, Yi and Lin, Yu and Lin, Yun},
  title   = {Music Audio Emotion Regression Using the Fusion of Convolutional Neural Networks and Bidirectional Long Short-Term Memory Models},
  journal = {PeerJ Comput. Sci.},
  volume  = {11},
  pages   = {e3086},
  year    = {2025},
  doi     = {10.7717/peerj-cs.3086}
}

@inproceedings{multimodal2018,
  author    = {Delbouys, R{\'e}mi and others},
  title     = {Music Mood Detection Based on Audio and Lyrics with Deep Neural Net},
  booktitle = {Proc. ISMIR},
  year      = {2018}
}

@inproceedings{llm_annotation2025,
  author    = {Yang, Meng and others},
  title     = {Exploring the Feasibility of {LLMs} for Automated Music Emotion Annotation},
  booktitle = {Proc. ISMIR},
  year      = {2025}
}

@inproceedings{manyshot_icl2024,
  author    = {Agarwal, Rishabh and others},
  title     = {Many-Shot In-Context Learning},
  booktitle = {Proc. NeurIPS},
  volume    = {37},
  pages     = {76930--76966},
  year      = {2024}
}

@inproceedings{brown2020gpt3,
  author    = {Brown, Tom B. and others},
  title     = {Language Models Are Few-Shot Learners},
  booktitle = {Proc. NeurIPS},
  volume    = {33},
  year      = {2020}
}

@book{R2_1998,
  author    = {Draper, Norman R. and Smith, Harry},
  title     = {Applied Regression Analysis},
  publisher = {McGraw-Hill},
  year      = {1998}
}

@article{BEATs2022,
  author  = {Chen, Sanyuan and others},
  title   = {{BEATs}: Audio Pre-Training with Acoustic Tokenizers},
  journal = {arXiv preprint arXiv:2212.09058},
  year    = {2022}
}

@inproceedings{li2024mert,
  author    = {Li, Yizhi and others},
  title     = {{MERT}: Acoustic Music Understanding Model with Large-Scale Self-Supervised Training},
  booktitle = {Proc. ICLR},
  year      = {2024}
}

@article{qwen3omni2025,
  author  = {Xu, Jin and others},
  title   = {{Qwen3-Omni} Technical Report},
  journal = {arXiv preprint arXiv:2509.17765},
  year    = {2025},
  doi     = {10.48550/arXiv.2509.17765}
}

@article{qwen2audio,
  author  = {Chu, Yunfei and others},
  title   = {{Qwen-Audio}: Advancing Universal Audio Understanding via Unified Large-Scale Audio-Language Models},
  journal = {arXiv preprint arXiv:2311.07919},
  year    = {2023}
}

@article{defossez2021hybrid,
  author  = {D{\'e}fossez, Alexandre},
  title   = {Hybrid Spectrogram and Waveform Source Separation},
  journal = {arXiv preprint arXiv:2111.03600},
  year    = {2021}
}

@inproceedings{radford2023whisper,
  author    = {Radford, Alec and others},
  title     = {Robust Speech Recognition via Large-Scale Weak Supervision},
  booktitle = {Proc. ICML},
  year      = {2023}
}

\end{document}